\documentclass[a4paper, submission, Proceedings]{SciPost}

\hypersetup{
    colorlinks,
    linkcolor={red!50!black},
    citecolor={blue!50!black},
    urlcolor={blue!80!black}
}

\usepackage[bitstream-charter]{mathdesign}
\usepackage{url}
\usepackage{hyperref}
\usepackage{physics}
\usepackage[compat=1.1.0]{tikz-feynman} 
\usepackage[english]{babel}
\usepackage{graphicx}
\usepackage{float}
\usepackage{xcolor}
\usepackage{caption}
\usepackage{subcaption}
\usepackage[utf8]{inputenc}
\usepackage{enumerate}
\usepackage{multirow}
\usepackage{dsfont}
\usepackage{slashed}
\usepackage{textcomp}
\usepackage{cite}
\usepackage{authblk}
\usepackage{lineno}

\DeclareSymbolFont{usualmathcal}{OMS}{cmsy}{m}{n}
\DeclareSymbolFontAlphabet{\mathcal}{usualmathcal}

\begin{document}

\begin{center}{\Large \textbf{
Perturbative heavy quark contributions to the anomalous magnetic moment of the muon\\
}}\end{center}

\begin{center}
P. D. Kennedy\textsuperscript{1*},
J. Erler\textsuperscript{2} and
H. Spiesberger\textsuperscript{1}
\end{center}

\begin{center}
{\bf 1} PRISMA$^+$ Cluster of Excellence, 
Institute of Physics, Johannes Gutenberg Universit\"{a}t, D-55099 Mainz, Germany
\\
{\bf 2} PRISMA$^+$ Cluster of Excellence, 
Institute of Nuclear Physics, Johannes Gutenberg Universit\"{a}t, D-55099 Mainz, Germany
\\
* philip.david.kennedy@cern.ch
\end{center}

\begin{center}
\today
\end{center}


\definecolor{palegray}{gray}{0.95}
\begin{center}
\colorbox{palegray}{
  \begin{minipage}{0.95\textwidth}
    \begin{center}
    {\it  16th International Workshop on Tau Lepton Physics (TAU2021),}\\
    {\it September 27 – October 1, 2021} \\
    \doi{10.21468/SciPostPhysProc.?}\\
    \end{center}
  \end{minipage}
}
\end{center}

\section*{Abstract}
{\bf
We discuss a method for calculating the heavy quark vacuum polarisation contribution to the muon anomalous magnetic moment, $a_\mu$, using perturbative QCD. This approach is independent of $e^+e^-$ cross-section data allowing a fully theoretical evaluation of these contributions. We confirm an existing result at lower orders in $\alpha_s$ and state a new explicit analytic formula which includes terms up to $\order{\alpha_s^3}$. Numerically the charm quark contribution to $a_\mu$ is found to be $a_\mu^c = (14.5 \pm 0.2)\times 10^{-10}$ and the bottom contributes $a_\mu^b = (0.302 \pm 0.002) \times 10^{-10} $. Our uncertainty estimates include both parametric uncertainties from $\hat{m}_q(\hat{m}_q)$ and $\alpha_s(\hat{m}_q)$, and theoretical uncertainties in the perturbative expansion. Comparison is made between these results and alternative approaches such as lattice QCD or those based on a dispersion relation and cross-section data.
}

\vspace{10pt}
\noindent\rule{\textwidth}{1pt}
\tableofcontents\thispagestyle{fancy}
\noindent\rule{\textwidth}{1pt}
\vspace{10pt}

\section{Introduction}
\label{sec:intro}
This work builds upon the previous work of Erler and Luo who  calculated QCD components of the hadronic vacuum polarisation (HVP) contributions to the anomalous magnetic moment of the muon \cite{Erler:2000nx}. This evaluation took place 20 years ago using QCD input up to $\order{\alpha_s^2}$. Since then the necessary vector current correlator has been expanded up to $\order{\alpha_s^3}$ and the precision of both the $\overline{\text{MS}}$-scheme heavy quark masses $\hat{m}_q(\hat{m}_q)$, and the strong coupling constant $\alpha_s(M_Z)$ have improved significantly. It is therefore timely to carry out a new evaluation of the heavy quark contributions to the anomalous magnetic moment of the muon, $a_\mu^q$. 
\par 
Our method will allow us to state the heavy quark contributions as both an explicit formula for $a_\mu^q$, valid for all heavy quarks, and numerical evaluations of this formula. The explicit formula allows adjustment to be made for different values of the $\overline{\text{MS}}$-scheme heavy quark masses and the strong coupling constant. It is also possible to apply this formula for heavy quark contributions to other lepton magnetic moments. In the case of the electron, where experimental precision is improving rapidly~\cite{Parker:2018vye,Morel:2020dww}, it will now be pertinent to state these contributions. However, in the case of the tau lepton experimental precision is currently far below the order of magnitude at which heavy quarks would contribute.
\par 
In \autoref{sec:method} we will outline the methods used to carry out this calculation. The results for $a_\mu^q$ as both an explicit formula and numerical results will be shown in \autoref{sec:results}. These numerical results will then be compared to existing results from both perturbative QCD (pQCD) and a selection of LQCD evaluations in \autoref{sec:comparison}. An aside will then be given for the electron anomalous magnetic moment where we state the numerical results for these heavy quark contributions in \autoref{sec:electron} before concluding in \autoref{sec:conclusion}.

\section{Method}\label{sec:method}
\subsection{Outline of hadronic contributions}
\begin{figure}[t!]
	\begin{subfigure}{0.5\textwidth}
			\begin{center}	
			\begin{tikzpicture}
				\begin{feynman}
					\vertex (v);
					\vertex[below left=1.5cm of v] (r1);
					\vertex[right=0.7cm of r1] (l1);
					\vertex[below left=2.5cm of v] (i) {\(\mu\)};
					\vertex[below right=1.5cm of v] (r2);
					\vertex[left=0.7cm of r2] (l2);
					\vertex[below right=2.5cm of v] (f) {\(\mu\)};
					\vertex[above = 1cm of v] (p);
					
					\diagram* {
						{[edges=fermion]
							(i) -- (r1) -- (v) -- (r2) -- (f),
						},
						(v) -- [boson, edge label=\(\gamma\)] (p),
						(l1) -- [boson, edge label=\(\gamma\)] (r1),
						(r2) -- [boson, edge label=\(\gamma\)] (l2),
						(l1) -- [fermion, red, out=90, in=90, edge label=\(q\)] (l2),
						(l2) -- [fermion, red, out=-90, in=-90, edge label=\(\bar{q}\)] (l1),
					};
				\end{feynman}
			\end{tikzpicture}
		\end{center} 
	\end{subfigure}
	\begin{subfigure}{0.5\textwidth}
			\begin{center} 
			\includegraphics[scale=0.25]{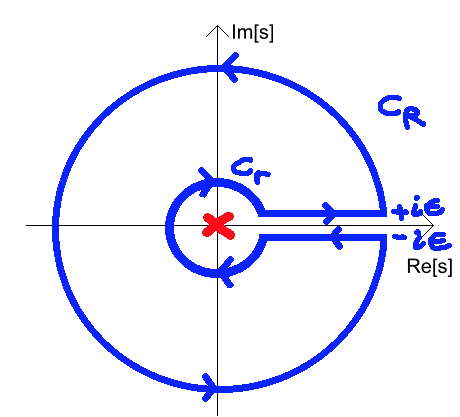}
			\end{center}
	\end{subfigure}
\caption{(a) The leading order hadronic vacuum polarisation contribution to $a_\mu^q$. 
In pQCD it is described by a quark-loop insertion, shown in red. 
(b) Contour in the complex plane showing the application of Cauchy's theorem around a singularity at the origin.}
			\label{fig:contour}
\end{figure}

Hadronic contributions to the anomalous magnetic moment of the muon are currently the dominating source of theoretical error within the standard model. As a result, this currently attracts significant interest and there are multiple groups working on evaluations of these contributions. The leading order Feynman diagram can be seen in the left of \autoref{fig:contour} for the case of quark contributions. The primary formula used to evaluate it is
\begin{equation}
	\eval{a_\mu^\text{had}}_{2\text{-loop}} = \bigg(\frac{\alpha m_\mu}{3\pi}\bigg)^2 \int_{s_{th}}^{\infty} \frac{ds}{s^2}\hat{K}(s)\cdot R(s) 
	\, , 
	\hspace{40pt}
	R(s) = 12\pi \Im{{\Pi(s+i\epsilon)}} 
\label{eq:2-loop_integral}
\end{equation}
where
\begin{equation}
	\hat{K}(s) 
	= 
	\int_{0}^{1}dx 
	\frac{3x^2(1-x)}{\frac{m_\mu^2}{s}\cdot x^2 + (1-x)}\ .
		\label{eq:Kernal_integral-1}
\end{equation}
\autoref{eq:2-loop_integral} is commonly calculated using a numerical form of the kernel, in \autoref{eq:Kernal_integral-1}, and $e^+ e^-$ cross-section data\cite{Davier:2019can, Keshavarzi:2019abf} for 
$$ R(s) = \frac{\sigma(e^+e^- \rightarrow \text{hadrons})}{\sigma(e^+e^-\rightarrow \mu^+\mu^-)}\ . $$
Our aim is to use pQCD in order to make our calculation for heavy quarks independent of data. \par 

\subsection{Our method}
Knowledge of $R(s)$ within pQCD is limited to 3 main ranges $s\rightarrow0$, $s = s_{th.}$ and $s\rightarrow\infty$, where $s_{th.}$ is a given threshold. Within these ranges specific expansions are known up to $\order{\alpha_s^3}$. However, in between these regions the knowledge of $R(s)$ has limited numerical precision. It is therefore favourable to use a method which can evaluate $a_\mu^q$ within one of these ranges to avoid the reduced precision outside these areas. Our method will use Cauchy's theorem to transform \autoref{eq:2-loop_integral} from an integral along the real axis to a contour integral in the complex plane at some fixed radius $s=s_0$. Cauchy's theorem for an analytically continuous function $f(s)$, 
\begin{equation}
	\oint ds \, f(s) = 0 \, , 
	\label{eq:cauchy-theorem}
\end{equation}
is used to derive the relation 
\begin{equation}
	2i \lim_{R \to \infty} 
	\int_{s_0}^R \frac{ds}{s^2} \hat{K}(s) \Im{\Pi(s)} 
	= 
	- \int_{C_r} \frac{ds}{s^2} \hat{K}(s) \Pi(s)  
	\label{eq:cauchy-theorem-transform}
\end{equation}
which can be used to calculate the hadronic contribution to $a_\mu$: 
\begin{equation}
	a_\mu^\text{had} = 
	\frac{1}{2i}
	\left(\frac{\alpha m_\mu}{3\pi}\right)^2 12\pi 
	\int_{\bar{C}_r} \frac{ds}{s^2} \hat{K}(s) \Pi(s) \, .
	\label{eq:master_transform}
\end{equation}
The integration contour is illustrated by the right of \autoref{fig:contour}. $\bar{C}_r$ is the anti-clockwise contour at radius $r$.  

The kernel in \autoref{eq:Kernal_integral-1} is not in a convenient form for this calculation as a double integral arising from \autoref{eq:master_transform} would make an explicit analytic evaluation extremely difficult. Therefore, it is favourable to  expand \autoref{eq:Kernal_integral-1} into a more convenient form for the energy range in which we will evaluate this. The initial high-energy expansion chosen in Ref.~\cite{Erler:2000nx} is
\begin{equation}
	\eval{\hat{K}(s)}_{\text{Exp.}} = 1 + \bigg(\frac{m_\mu^2}{s}\bigg)\cdot\bigg[\frac{25}{4}+3\ln(\frac{m_\mu^2}{s})\bigg] + \order{\frac{m_\mu^4}{s^2}}.
	\label{eq:kernel_expansion}
\end{equation}
This simplifies the calculation while remaining valid for the energy regime we are investigating. The drawback of this expansion is that the logarithmic term in \autoref{eq:kernel_expansion} is not analytically continuous and therefore an alternative method to Cauchy's theorem will be needed for integration of this term. 
To solve this issue we divide the kernel in \autoref{eq:kernel_expansion} into three separate pieces which will be integrated separately. The separation used here is
\begin{equation}
	\eval{\hat{K}(s)}_{\text{Exp.}} = 1 + \frac{m_\mu^2}{s}\left(\frac{25}{4} + 3\ln(\frac{s_0}{s})\right) + \frac{m_\mu^2}{s}\left( 3\ln(\frac{m_\mu^2}{s_0})\right).
	\label{eq:kernel-division}
\end{equation}
This expansion is chosen as it separates the term containing $\ln(s)$ for which Cauchy's theorem cannot be used. It will also allow us to perform a direct comparison with the original explicit result \cite{Erler:2000nx}. For a convenient comparison the introduced constant $s_0$ will be set equal to the heavy quark mass $\hat{m}_q$ in the $\overline{\text{MS}}$-scheme.
\par 
It is now possible to perform the contour integral in \autoref{eq:master_transform} for the terms without $\ln(s)$ using the well-known expression of the heavy quark vector current correlator up to $\order{\alpha_s^3}$ from Refs.~\cite{Maier:2009fz, Hoang:2008qy}. The integral is performed at a fixed radius of $r=\hat{m}_q$. For the term containing $\ln(s)$ the integration must be performed along the real axis using \autoref{eq:2-loop_integral}. This will involve using both the threshold and high-energy expansions for the vector current correlator at $\order{\alpha_s^2}$ and $\order{\alpha_s^3}$\cite{Chetyrkin:1996cf,Chetyrkin:2006xg} interpolating between them. We will use a spread of eight different interpolation methods to calculate an extremely conservative error on this term. However, as we will see in \autoref{sec:results}, these uncertainties are on terms which have very small contributions to the overall result for $a_\mu^q$; therefore this approach will not limit the final precision. At $\order{\alpha_s^0}$ and $\order{\alpha_s^1}$ the full explicit form of $\Im{\Pi(s)}$ is known and can be calculated exactly.

\begin{table}[t!]
	\begin{center}
		\begin{tabular}{|c||c|c|}
			\hline
			Quark & Charm & Bottom  \\
			\hline\hline
			$\Delta \hat{m}_q$ & $-0.18$ & $-0.0011$ \\
			\hline
			$\Delta \alpha_s(\hat{m}_q)$ & $0.19$ & $0.0013$ \\
			\hline
			(Anti-)Correlation with $\Delta\alpha_s(\hat{m}_q)$ & $-0.09$ & $0.0001$ \\
			\hline\hline
			Total & $0.21$ & $0.0018$ \\
			\hline
		\end{tabular}
		\caption{The error induced on $a_\mu^q$ (in units of $10^{-10}$) due to the uncertainty on the input parameters. The correlation between these two parameters is also calculated. }
		\label{tab:mu_budget}
	\end{center}
\end{table}

\section{Results}\label{sec:results}
Following the integration of all three terms of the kernel in \autoref{eq:kernel-division} separately, it is now possible to state an explicit form for $a_\mu^q$. This is found to be:
\begin{eqnarray}
	a_\mu^{q} 
	&=& 
	\frac{\alpha^2 Q_q^2}{4\pi^2}
	\left[
	\frac{m_\mu^2}{4\hat{m}_q^2}
	\left(\frac{16}{15} + \frac{3104}{1215}a_s \right. 
	\right. 
	\label{eq:explicit-result}
	\\  
	&& \qquad\qquad\qquad \left.
	+ \left(0.50988 + \frac{2414}{3645}n_l\right)a_s^2 
	+ \left(1.87882 - 2.79492n_l + 0.09610n_l^2\right)a_s^3
	\right)
	\nonumber \\
	&& \qquad\quad
	+ \frac{m_\mu^4}{16\hat{m}_q^4}
	\left(\frac{108}{1225}
	- 0.194294 \cdot a_s
	+ \left(-15\pm28 -\left(1\mp1 \right)n_l \right) \cdot a_s^2
	\right)  
	\nonumber \\[1ex] 
	&& \qquad\quad 
	+ 3 \frac{m_\mu^4}{16\hat{m}_q^4} 
	\ln(\frac{m_\mu^2}{\hat{m}_q^2})
	\Biggl(\frac{16}{35} + \frac{15728}{14175}a_s   
	\nonumber \\
	&& \qquad\qquad\qquad\qquad\qquad\qquad 
	+ \left(1.41227 + \frac{290179}{637875}n_l\right)a_s^2 
	\nonumber \\ 
	&&  \qquad\qquad\qquad\qquad\qquad\qquad
	\left. \left.
	+\left(-6.23488 + 0.96156n_l -0.01594n_l^2\right)a_s^3
	\right)
	\Biggr)
	\right] 
	\nonumber
\end{eqnarray}
where $\alpha$ is the fine structure constant, $Q_q$ is the charge of the heavy quark, $a_s = \alpha_s(\hat{m}_q)/\pi$, and $n_l = n_f-1$ is the number of light quarks. This expression is valid for all heavy quarks. The uncertainties given in this expression are calculated as the range of different methods used to interpolate between different energy regimes as explained in \autoref{sec:method}.

It is possible to compare this result to that of Ref.~\cite{Erler:2000nx}, which was calculated up to $\order{\alpha_s^2}$. We find exact agreement in all comparable terms and we have improved the precision on all numerical terms by two significant figures. In addition we now include terms up to $\order{\alpha_s^3}$.

Having evaluated $a_\mu^q$ explicitly we can now evaluate our result for the two relevant heavy quarks, charm and bottom. 
To allow for a meaningful comparison of pQCD with LQCD we will use input values which are themselves independent of LQCD. 
The quark masses are taken from Refs.~\cite{Erler:2016atg, Masjuan:2021}, 
\begin{eqnarray}
	\hat{m}_c(\hat{m}_c) &=& 1.273 \pm 0.009 \text{GeV} 
	\quad \text{for} \quad \alpha_s(M_Z) = 0.1185\pm0.0016 \, , 
	\\
	\hat{m}_b(\hat{m}_b) &=& 4.180 \pm 0.008 \text{GeV} 
	\quad \text{for} \quad \alpha_s(M_Z) = 0.1185\pm0.0016 \, . 
\end{eqnarray}
The values of the strong coupling constant at the quark masses $~\alpha_s(\hat{m}_q)$ are
\begin{eqnarray}
	\alpha_s(\hat{m}_c) = 0.396 \pm 0.020 \, , 
	&\space&
	\alpha_s(\hat{m}_b) = 0.2267 \pm 0.0061 \, , 
\end{eqnarray}
given $\alpha_s(M_Z) = 0.1185\pm0.0016$ in the EW fit from Ref.~\cite{Zyla:2020zbs}. The running of $\alpha_s$ has been calculated using CRunDec~\cite{Schmidt:2012az}.

\begin{table}[t!]
	\begin{center}
		\begin{tabular}{|c||c|c|c|c|c|}
			\hline
			$a_\mu^c \cdot 10^{10}$ & $\order{\alpha_s^0}$ & $\order{\alpha_s^1}$ & $\order{\alpha_s^2}$ & $\order{\alpha_s^3}$ & $\order{\alpha_s^4}$ \\
			\hline\hline
			\color{red}$\order{\frac{m_\mu^2}{\hat{m}_q^2}}$ & \color{red}11.0131 & \color{red}3.3214 & \color{red}0.4087 & \color{red}0.1163 & $<\pm 0.12$ \\
			\hline\hline
			\color{purple}$\order{\frac{m_\mu^4}{\hat{m}_q^4}\ln(\frac{m_\mu^2}{\hat{m}_q^2})}$ & \color{purple}$-0.1214$ & \color{purple}$-0.0371$ & \color{purple}$-0.0117$ & \color{purple}0.0019 & $<\pm 0.002$ \\
			\hline
			\color{blue}$\order{\frac{m_\mu^4}{\hat{m}_q^4}}$ & \color{blue}0.0016 & \color{blue}$-0.00044$ & \color{blue}$-0.0034$ & \color{blue}$0.00004A_{32}^0$ & $<\pm0.00004A_{32}^0$ \\
			\hline\hline
			\color{orange}$\order{\frac{m_\mu^6}{\hat{m}_q^6}\ln(\frac{m_\mu^2}{\hat{m}_q^2})}$ & \color{orange}$-0.00074$ & \color{orange}$-0.00018$ & \color{orange}$-0.000072$ & \color{orange}0.000016 & $<\pm0.00002$ \\
			\hline
			\color{orange}$\order{\frac{m_\mu^6}{\hat{m}_q^6}}$ & \color{orange}$-0.000031$ & \color{orange}$-0.000020$ & \color{orange}$5A_{23}^0\cdot10^{-7}$ & \color{orange}$6A_{33}^0\cdot10^{-8}$ &  $<\pm6A_{33}^0\cdot10^{-8}$\\
			\hline
		\end{tabular}
		\caption{Numerical evaluation of individual terms in \autoref{eq:explicit-result} for the case of the charm quark. It includes estimated bounds on as yet unknown terms and calculation of higher-order terms from the kernel (yellow). All values are in units of $10^{-10}$.}
		\label{tab:charm-coefficients}
	\end{center}
	\begin{center}
		\begin{tabular}{|c||c|c|c|c|c|}
			\hline
			$a_\mu^b \cdot 10^{10}$ & $\order{\alpha_s^0}$ & $\order{\alpha_s^1}$ & $\order{\alpha_s^2}$ & $\order{\alpha_s^3}$ & $\order{\alpha_s^4}$ \\
			\hline\hline
			\color{red}$\order{\frac{m_\mu^2}{\hat{m}_q^2}}$ & \color{red}0.255335 & \color{red}0.044128 & \color{red}0.003937 & \color{red}$-0.000698$ & $<\pm 0.0007$ \\
			\hline\hline
			\color{purple}$\order{\frac{m_\mu^4}{\hat{m}_q^4}\ln(\frac{m_\mu^2}{\hat{m}_q^2})}$ & \color{purple}$-0.00039$ & \color{purple}$-0.000068$ & \color{purple}$-0.000014$ & \color{purple}0.0000008 & $<\pm0.0000008$ \\
			\hline
			\color{blue}$\order{\frac{m_\mu^4}{\hat{m}_q^4}}$ & \color{blue}0.000003 & \color{blue}$-5\cdot10^{-7}$ & \color{blue}$-0.000002$ & \color{blue}$A_{32}^0\cdot10^{-8}$ & $<\pm A_{32}^0\cdot10^{-8}$ \\
			\hline\hline
			\color{orange}$\order{\frac{m_\mu^6}{\hat{m}_q^6}\ln(\frac{m_\mu^2}{\hat{m}_q^2})}$ & \color{orange}$-2\cdot10^{-7}$ & \color{orange}$-3\cdot10^{-8}$ & \color{orange}$-9\cdot 10^{-9}$ & \color{orange}$4\cdot 10^{-10}$ & $<\pm 4\cdot 10^{-10}$ \\
			\hline
			\color{orange}$\order{\frac{m_\mu^6}{\hat{m}_q^6}}$ & \color{orange}$-6\cdot10^{-9}$ & \color{orange}$-2\cdot10^{-9}$ & \color{orange}$ 3A_{23}^0 \cdot 10^{-11}$ & \color{orange}$2A_{33}^0 \cdot 10^{-12}$ & $<\pm 2A_{33}^0 \cdot 10^{-12} $ \\
			\hline
		\end{tabular}
		\caption{Same as \autoref{tab:charm-coefficients} for the case of the bottom quark.}
		\label{tab:bottom-coefficients}
	\end{center}
\end{table}

Using these values we find that
\begin{eqnarray}
	a_\mu^c = 14.5 \pm 0.2 \, , 
	&\hspace{60pt}&
	a_\mu^b = 0.302 \pm 0.002 \, , 
\end{eqnarray}
in units of $10^{-10}$. The parametric error budget induced in $a_\mu^q$ for both quark masses is displayed in \autoref{tab:mu_budget}. In the calculation we have taken into account the correlation between $\hat{m}_q$ and $\alpha_s(\hat{m}_q)$. In the case of the charm quark the correlation leads to a reduction of the overall error, while for the bottom quark we observe an enhancement.

In addition to parametric uncertainties we must also consider the theoretical uncertainties which are arising from the perturbative expansion of $\Pi(s)$ and the expansion of $\hat{K}(s)$. Our method for estimating these was to use our known lower order terms to act as an approximate upper bound for the value at higher orders. The results of evaluating all known terms and estimating bounds for higher order terms are shown in \autoref{tab:charm-coefficients} for the charm quark and \autoref{tab:bottom-coefficients} for the bottom quark. These tables show that the dominant contributions are likely to come from the $\order{\alpha_s^4 \frac{m_\mu^2}{\hat{m}_q^2}}$ terms which we estimate by using the $\order{\alpha_s^3}$ term as an upper bound. In both charm and bottom quark cases the largest contributions at higher orders are below that of the parametric uncertainties. Therefore, at the current parametric precision we do not consider theoretical uncertainties to be relevant to the overall total uncertainty.

\section{Comparison}\label{sec:comparison}
\begin{figure}[t!]
	\begin{subfigure}{0.49\textwidth}
		\begin{center}
			\includegraphics[scale=0.45]{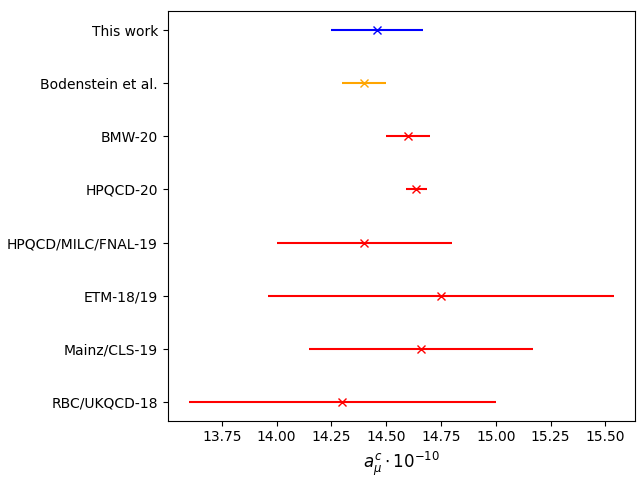}
		\end{center}
	\end{subfigure}
	\begin{subfigure}{0.49\textwidth}
		\begin{center}
			\includegraphics[scale=0.45]{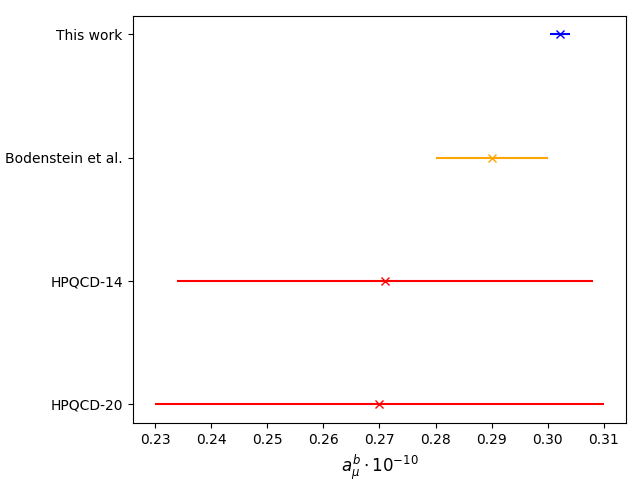}
		\end{center}
	\end{subfigure}
	\caption{Comparison of our results for (a) the charm quark contribution and (b) the bottom quark contribution 
	with a selection of results from the literature. 
	pQCD results are shown in yellow and LQCD results in red.}
	\label{fig:comparisons}
\end{figure}

In \autoref{fig:comparisons} our results for $a_\mu^c$ and $a_\mu^b$ are compared with a selection of recent results from both pQCD and LQCD. Both of our results agree with these other results within $1\sigma$. For the case of the charm quark, shown in the left panel of \autoref{fig:comparisons}, our result is in very good agreement with all LQCD and pQCD values. The point labelled Bodenstein et al.~\cite{Bodenstein:2011qy} is currently the only other published result from a pQCD approach and uses a numerically fitted kernel. This research was repeated during this work and we found that neither the parametric uncertainty from $\alpha_s(\hat{m}_q)$ nor its correlation with $\hat{m}_q$ had been included within the error calculation. As a result when this was repeated using our method of uncertainty propagation we found an error approximately twice the published one and thus larger than our own.

For the case of the bottom quark (right panel of \autoref{fig:comparisons}) our result is in good agreement with all other points within $1\sigma$. There is a $1\sigma$ tension with Bodenstein et al.\cite{Bodenstein:2011qy} but, as for the charm quark, this error appears to be underestimated by approximately a factor of two due to the absence of the $\alpha_s(\hat{m}_b)$ parametric uncertainty and its correlation with $\hat{m}_b$. If the uncertainty approach were similar to ours there would be extremely good agreement as was the case for the charm. Our result for $a_\mu^b$ has the highest precision compared with other results available in the literature at the time of writing. In the case of the bottom quark we find a particularly high precision due to its large mass reducing the induced effect on the overall uncertainty. 

\section{$(g-2)_e$}\label{sec:electron}
As the mass of the electron is below that of the muon our result in \autoref{eq:explicit-result} can be applied to heavy quark effects for $a_e^q$ without any loss in validity. Our formula allows quick and convenient re-calculation for the electron by simply replacing the muon mass. We find the results 
\begin{eqnarray}
	a_e^c = 34.2 \pm 0.5 \, , 
	&\hspace{40pt}&
	a_e^b = 0.708 \pm 0.004 \, , 
\end{eqnarray}
in units of $10^{-15}$. The same method for calculating the parametric uncertainty was carried out here and the results are displayed in \autoref{tab:electron_parametric}. As in the case of the muon there is a significant anti-correlation between $\alpha_s(\hat{m}_q)$ and $\hat{m}_q$ for the charm quark which reduces the overall uncertainty. For the bottom quark there is a small correlation increasing the overall uncertainty. A calculation of the theoretical uncertainties similar to \autoref{tab:charm-coefficients} and \autoref{tab:bottom-coefficients} has been carried out and the largest contributions are found to be below the parametric uncertainties.

\section{Conclusion}\label{sec:conclusion}
In the course of this research we have reproduced an existing result for the heavy quark contribution to the anomalous magnetic moment of the muon $a_\mu^q$ in pQCD\cite{Erler:2000nx}. Our approach is based on an expansion of the integration kernel in the formula for $a_\mu$ that allows us to apply Cauchy's theorem. The radius for the integration contour can then be chosen in a range where pQCD is applicable. 
We have successfully improved the precision of all numerical coefficients and expanded the formula to $\order{\alpha_s^3}$. This has allowed us to produce a highly precise result for $a_\mu^c = (14.5 \pm 0.2)\times 10^{-10}$ which is independent from LQCD. Our result for $a_\mu^b = (0.302 \pm 0.002) \times 10^{-10} $ is the most precise result in the literature at the time of writing. Our two results demonstrate good agreement between pQCD and LQCD. Future improvements in LQCD specifically for the case of the bottom quark would act as a further precision test between the two methods. Our explicit formula for $a_\mu^q$ allowed for calculation of the numerical results for $a_e^q$ and will allow for further improvement in the numerical results for $a_\mu^q$ following future improvement in $\hat{m}_q$ and $\alpha_s(\hat{m}_q)$.

\begin{table}[t!]
	\begin{center}
		\begin{tabular}{|c||c|c|}
			\hline
			Quark & Charm & Bottom  \\
			\hline\hline
			$\Delta \hat{m}_q$ & $-0.43$ & $-0.0027$ \\
			\hline
			$\Delta \alpha_s(\hat{m}_q)$ & $0.46$ & $0.0031$ \\
			\hline
			(Anti-)Correlation with $\Delta\alpha_s(\hat{m}_q)$ & $-0.22$ & $0.0002$ \\
			\hline\hline
			Total & $0.49$ & $0.0042$ \\
			\hline
		\end{tabular}
		\caption{The error (in units of $10^{-15}$) induced in $a_\mu^e$ due to the uncertainty in the input parameters. 
		The correlation between these two parameters is also calculated.}
		\label{tab:electron_parametric}
	\end{center}
\end{table}

\bibliography{mybib.bib}


\end{document}